# TODAY AND FUTURE NEUTRINO EXPERIMENTS AT KRASNOYARSK NUCLEAR REACTOR

Yu.V.Kozlov[a], S.V.Khalturstsev[a], I.N.Machulin, A.V.Martemyanov[a], V.P.Martemyanov[a], A.A.Sabelnikov[a], V.G.Tarasenkov[a], E.V.Turbin[a], V.N.Vyrodov[a], L.A.Popeko[b], A.V.Cherny[b], G.A.Shishkina[b]

[a]Russian Research Center "Kurchatov Institute" 123182 Moscow, Russia
[b]St-Petersburg Nuclear Physics Institute 188350, Gatchina, St-Petersburg district, Russia

The results of undergoing experiments and new experiment propositions at Krasnoyarsk underground nuclear reactor are presented.

## 1. KRASNOYARSK UNDERGROUND NEUTRINO LABORATORY

The underground neutrino laboratory of RRC "Kurchatov Institute" is situated near the Krasnoyarsk nuclear reactor at Sayan hills by the Enisey river in the depth of about 600 meters of the water equivalent (m.w.e.). The laboratory consists of two experimental halls. One of them (the hall $N°1$) is placed ~ 30m far from the reactor core and another one (the hall $N°2$) ~ 20m far from the center of the reactor. The experiments on the investigation of the reactor antineutrinos interactions with protons, deuterons, and electrons were carried out since 1982 in the hall $N°1$.

## 2. DEUTERON DISINTEGRATION INDUCED BY THE REACTOR ANTINEUTRINOS

There are two channels of the antineutrino-deuteron interaction:
the Neutral Current disintegration of the Deuteron

(NCD): $\tilde{\nu}_e + d \rightarrow \tilde{\nu}'_e + p + n$,

and the Charged Current disintegration of the Deuteron

(CCD): $\tilde{\nu}_e + d \rightarrow e^+ + n + n$.

The investigation of these reactions can give information about (a) the weak-interaction constants of Charged and Neutral Currents, (b) the neutrino oscillations, and (c) the scattering length for the neutron–neutron interaction. The investigation of the reactor antineutrino interaction with deuteron is continued at the Krasnoyarsk neutrino laboratory with the aim to improve the accuracy of the measured cross sections for the disintegration of deuteron by low energy electron antineutrinos through Neutral (NCD) and Charged-Current (CCD) channels. The previous experiments on the study of the antineutrino–deuteron interaction gave precision for the cross sections 12% [1]. A new Krasnoyarsk experiment is performed to improve the precision up to 8%. The detector target consist of stainless steel tank $90\times90\times70$ cm$^3$, containing 513 liters of $D_2O$. More detailed description of the detector is given in [2]. An efficiency of the detector for the neutrons induced by antineutrino-deuteron interactions and the $^{252}$Cf-neutron source were calculated by the Monte-Carlo simulation. The comparison of the calculations with the experimental results for the $^{252}$Cf-neutron source showed that the difference is less than 1%. The efficiency of the one neutron registration by the tank's counters is (56.0±0.7)%, and of the two neutron registration by all of counters is (41.6±0.4)%. The neutron lifetime in the detector is (203±2)$\mu sec$

Special attention was given to the correlated background for the NCD channel connected with the antineutrino–proton interaction, because of the cross section for this process is one order of magnitude greater than the antineutrino–deuteron interaction one. Therefore, the materials for the detector were chosen to minimize the content of hydrogen The construction of the detector was made to decrease that influence up to 0.002% by using the boron polyethylene. The concentration of the hydrogen in the heavy water is estimated as 0.15%. We expect that the correlated background would be 0.6 events per day.

The experiment is controlled by PC based CAMAC system. The collected information include:
a) the amplitude of the first and second neutrons;
b) the astronomic and life time;
c) the neutron registration zone number (the detector is divided on 24 registration zones);
d) the multiplicity of the events (the number of neutrons registered during 800$\mu sec$ time interval (the registration period of one event);
e) the information about the veto signal during the 800$\mu sec$ time interval before and after the neutron registration;
f) the time interval between neutron registrations inside the event registration period of the event.

The criteria for events acceptance are as follows:

a) the amplitude of the first neutron is in the energy interval should be in (644–884)keV;
b) the amplitude of the second neutron is in the energy interval should be in (190–884)keV;
c) the time between the first and the second neutron should be in the time interval (5-800)$\mu sec$;
d) no signals from the veto system during 800$\mu sec$ before and after neutron signal;
e) the multiplicity is less than 3 (no more then 2 neutrons are registered during 800 $\mu sec$).

Taking into account the real amplitude and the time selection criteria we had corrected the efficiencies: efficiency of the one neutron registration by the tank counters (NCD) is 50.7%, and efficiency of the two neutron registration by all counters (CCD) is 36.1%.

The data collection was started at the beginning of 1997 and is going on.

The results obtained during 360 days data collecting for reactor "ON" and 120 days for reactor "OFF" are: 24.39±2.24 events per day for the single neutron events, and 4.44±0.47 events per day for the two neutron events. To be sure that the electronic and the background conditions are stable, the analyses of the events with the neutron multiplicity more than 2 was made. As results for "ON"-"OFF" 0.02 ± 0.24 events per day have been obtained. They prove a high stability of the measurement conditions. After correction of the registration probability in NCD channel events corresponded CCD channel and taking into account the correlated background for inverse beta-decay on the protons from $H_2$ impurity. The value of (NCD)=18.3±1.71 events per day was obtained.

As a result of the experimental cross-sections we give the following values:

$${}^{NCD}\sigma_{\exp} = (3.09\pm0.30)\times10^{-44} cm^2/fis.\ {}^{235}U,$$

$${}^{CCD}\sigma_{\exp} = (1.05\pm0.12)\times10^{-44} cm^2/fis.\ {}^{235}U,$$

The theoretical predictions are:

$${}^{NCD}\sigma_{th} = (3.18\pm0.17)\times10^{-44} cm^2/fis.\ {}^{235}U,$$

$${}^{CCD}\sigma_{th} = (1.07\pm0.07)\times10^{-44} cm^2/fis.\ {}^{235}U$$

the corresponding ratios are as follows:

$${}^{NCD}\sigma_{\exp}/{}^{NCD}\sigma_{th} = 0.97\pm0.11,$$

$${}^{CCD}\sigma_{\exp}/{}^{CCD}\sigma_{th} = 0.99\pm0.13$$

## 3. SEARCH FOR NEUTRINO MAGNETIC MOMENT

A possible explanation of the solar neutrino problem (SNP) is assuming of the non-zero neutrino magnetic moment. This assumption may lead to the spin or the spin-flavor neutrino oscillations. Moreover, the suggestion of the magnetic interaction is following from a number of studies (Davis 1987, Oakley 1997, Rivin Yu.R. 1999)[3], which found non-correlation of the neutrino capture rate versus the indicators of the solar magnetic activity. Voloshin, Vysotski and Okun[4] have pointed that a magnetic moment of odder of magnitude $\sim10^{-11}\mu_B$ may explain a deficit of Solar neutrino on the Earth. That is why it is desirable to search for neutrino magnetic moment at this level. One of the most acceptable methods of the searching for neutrino moment is the measurement of the magnetic induced contribution to electron (anti) neutrino-electron scattering cross section for the low energy of recoil electrons. The contribution in the differential cross section of electron antineutrino– electron scattering due to magnetic moment is:

$$d\sigma^M/dT_e = (\pi\alpha^2\mu_v^2/m_e^2)(1-T_e/E_v)/T_e,$$

where $\mu_v$ is neutrino magnetic moment.
The corespondent contribution due to the weak interaction is:

$$d\sigma^W/dT_e = (G^2 m_e/2\pi)\times[(g_V+g_A)^2 +$$
$$(g_V-g_A)^2(1-E_e/E_v)^2 +$$
$$(m_e E_e/E_v^2)(g_V-g_A)^2]$$

where
$g_V = 1 + 2\sin^2\Theta_W$ and $g_A = -1/2$

Detector for these purposes must have the following properties:
a) high radioactive purity of detector materials;
b) low threshold for recoil electrons registration;
c) high energy resolution;
d) underground detector disposition (low cosmic correlated background.

A new antineutrino electron scattering measurement is planned to carry out at Krasnoyarsk underground neutrino laboratory (the hall 2) using the large silicon semiconductor detector developed and created in St.-Petersburg Nuclear Physics Institute [5].

The 80-kg silicon detector consists of four matrixes with 151 separate detectors in each. The detector placed in the vacuum nitrogen cooled cryostat in assembly with copper passive shield (of 90mm thickness). The cooled shielding, situated in vacuum, has 1600 kg of copper. It prevents the detector from Radon irradiation and from the electric contact system. HPGe detector 116cc of volume, 2 keV resolution is placed in the center of the silicon layout to control a gamma ray background. All of the detectors switched on in anti coincidence with the VETO system. VETO system consists of the 120 plastic scintillation detectors, surrounding the volume of the silicon detector. The electronic registration threshold is 50 keV. A trigger for the event is a appearance of only one silicon's signal. The

FADC digitized the sum signal from all of the detectors in the 100μsec interval before and after the triggered event.

This information is writing in the computer memory. The criteria of the pure ($\tilde{\nu}_e$,e)-event is a single pulse from one of the silicon modules within 200μsec time interval.

The direct measurements of the radioactivity of Si detector were performed in deep underground laboratory (Solotvino, Ukraine). As a passive shielding it was used pure semiconductor germanium. The detector was assembled in St. Petersburg Nuclear Institute and background measurements were carried out with and without passive shield. The results of the background measurements with expected effects from the reactor antineutrino - electron scattering at Krasnoyarsk neutrino laboratory are shown in the Fig. 1.

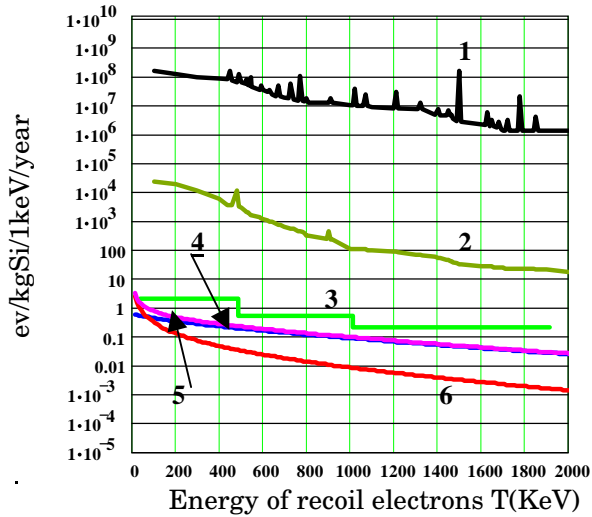

Figure1. The results of the background measurements.

1.-The background of the opened Ge detector at the PINP RAS laboratory (S-Petersburg);
2.-The background of the Ge detector covered by heavy shielding at the PINP RAS laboratory (S-Petersburg);
3.-The background of the 250g Si(Li) detector in Ge shielding- deep underground measurements (Solotvino - Ukraine);
4.-The spectrum of the scattered electrons due to antineutrino electron weak interaction plus magnetic moment $10^{-10}$ Bohr magnetons interaction for neutrino flux equal $7*10^{12}/cm^2*sec$
5.-The spectrum of scattered electrons due to the antineutrino electron weak interaction only for neutrino flux equal $7*10^{12}/cm^2*sec$
6.-The spectrum of the scattered electrons due to the antineutrino electron magnetic moment $5*10^{-11}$ Bohr magnetons interaction for neutrino flux equal $7*10^{12}/cm^2*sec$

Expected integral counting rate of the weak scattering is 100 per day. The expected background is about 1000 per day. It is planed to carry out reactor antineutrino-electron scattering experiment with Silicon semiconductor detector at Krasnoyarsk neutrino laboratory during 2000 -2002 with the aim of the searching for neutrino magnetic moment better then $5\times10^{-11}\mu_B$

**ACKNOWLEDGEMENTS**

The work is supported by Russian Foundation for Basic Research (grants ## 96-15-96610, 98-02-16313, 98-02-17614).

**REFERENCES**

1. Vyrodov.V.N. et al, JETPL,v.55, p.206(1992)
2. Yu.V.Kozlov, V.P.Martemyanov et al, JETP Lett., v.51, p.245 (1990);
3. Davis R.Proc.7$^{th}$ Work.on Grand Unif., ICOBAN"86, ed. R.Arafune (Singapore: World Scientific),273(1987). Oakley D.S., Snodgrass, Herschall B.,Astrop.Phys.,v.7, Issue 4,p.297(1997). Rivin Yu.R. et al e-print astro-ph/9902074(1999)
4. Voloshin M.B. at al, Zh Eks. Teor. Fiz., v.91, p754(1986)
5. Derbin A.V. et al, Pis'ma Zh. Eksp. Teor. Fiz., v.57,p755(1993)